\documentstyle[prb,aps,preprint]{revtex}

\newcommand{\navo}{$\alpha^\prime$-NaV$_2$O$_5$}

\begin{document}
\draft
\tighten
\title{One-dimensional dynamics of the $d$-electrons in \navo}
\author{S. Atzkern, M. Knupfer, M. S. Golden, J. Fink}
\address{Institute for Solid State Research, IFW Dresden, P.O.Box 270016, D-01171 Dresden, 
Germany}
\author{A. N. Yaresko \cite{imp}, V. N. Antonov \cite{imp}}
\address{Max-Planck-Institut f\"ur Physik komplexer Systeme, N\"othnitzer Str. 38, D-01187 
Dresden, Germany}
\author{A. H\"ubsch, C. Waidacher, K.~W.~Becker}
\address {Institut f\"ur Theoretische Physik, 
          Technische Universit\"at Dresden, D-01062 Dresden, Germany}
\author{W. von der Linden}
\address{Institut f\"ur Theoretische Physik, 
          Technische Universit\"at Graz, Petersgasse 16, A-8010 Graz, Austria}
\author{G. Obermeier, S. Horn}
\address{Institut f\"ur Physik, Universit\"at Augsburg, D-86159 Augsburg, Germany}
\date{\today}
\maketitle
\begin{abstract}
We have studied the electronic properties of the ladder compound \navo\, adopting a joint 
experimental and theoretical approach. The momentum-dependent loss function was measured 
using electron energy-loss spectroscopy in transmission. The optical conductivity derived from the 
loss function by a Kramers-Kronig analysis agrees well with our results from LSDA+$U$ band-
structure calculations upon application of an antiferromagnetic alignment of the V~3$d_{xy}$ spins 
along the legs and an on-site Coulomb interaction $U$ of between 2 and 3~eV. The decomposition 
of the calculated optical conductivity into contributions from transitions between selected energy 
regions of the DOS reveals the origin of the observed anisotropy of the optical conductivity. 
In addition, we have investigated the plasmon excitations related to transitions between the 
vanadium states within an effective 16 site vanadium cluster model. Good 
agreement between the theoretical and experimental loss function was obtained using the hopping 
parameters derived from the tight binding fit to the band-structure and moderate Coulomb 
interactions between the electrons within the $ab$ plane.
\end{abstract}

\pacs{71.10.Fd, 71.20.-b, 71.27.+a, 71.45.Gm, 78.20.-e}
\section{Introduction}
\label{introd}
In 1996 M. Isobe and Y. Ueda\cite{isobe} published magnetic susceptibility measurements of 
\navo\ powder samples, proposing the existence of linear antiferromagnetic spin-1/2 chains and a 
possible spin-Peierls transition at a critical temperature, T$_c$~=~34~K. The original picture of the 
charge ordering in this mixed valence ladder compound was that of alternating legs of V$^{4+}$ 
and V$^{5+}$ ions.\cite{carpy} However, a recent determination of the crystal structure using 
single crystal x-ray diffraction at room temperature\cite{smolinski,schnering} yielded only one 
symmetrically inequivalent vanadium (V$^{4.5+}$) position. Due to the observed one-dimensional 
character of the spin and charge system at room temperature - which was by that time also 
confirmed by inelastic neutron scattering\cite{fuji,yosihama} and angle-resolved photoemission 
experiments\cite{kobayashi} - the picture of linear chains of V-O-V rungs containing a single $d$-electron in a molecular orbital-like state with antiferromagnetic alignment along the ladders has 
increasingly won recognition. Below T$_c$ charge ordering has been observed in $^{51}$V NMR 
studies,\cite{ohama97} but the ordering pattern has not yet been defined. Calculations of the 
electronic ground state\cite{seo,mostovoy} have predicted a zig-zag charge ordering  along the 
ladders and only lately, with the support of the most recent experimental 
data\cite{konstantinovic,nakao,ohama00,lohmann00} does the evidence for zig-zag order appear 
to outweigh that for more complicated order patterns.\cite{luedecke} Nevertheless, there remain a 
number of important, open questions such as the nature of the correlations between the charges 
and spins which are the driving force for the observed charge ordering and which make the system 
insulating.

\par
The electronic structure of \navo\ has been intensively studied experimentally by means of optical 
absorption\cite{golubchik} and reflectivity,\cite{damascelli,long} as well as theoretically within the 
framework of bandstructure calculations \cite{smolinski,popovic,wu,katoh} and exact 
diagonalization techniques.\cite{cuoco,nishimoto98a,nishimoto98b} Nevertheless, there is a 
controversial discussion about the origin of the electronic excitations in this system. For example, 
concerning the absorption peak occurring at about 1~eV in optics, transitions between bonding 
and antibonding combinations of V~3$d_{xy}$ states,\cite{smolinski,damascelli,horsch} transitions 
between vanadium states of different symmetry\cite{long} as well as on-site $d-d$ transitions 
between crystal-field-split vanadium 3$d$ states\cite{golubchik} have been proposed as the origin 
of this feature. Furthermore, the information from the experiments still appears to be insufficient to 
enable the definition of a unique parameter set for the description of the electronic structure of 
\navo. In particular, independent of which theoretical model was used, significantly different values 
for the on-site Coulomb repulsion U, the inter-site Coulomb repulsion $V_{xy}$ between electrons 
on legs of adjacent ladders and the hopping parameter $t_{xy}$ have been discussed in the 
literature (for a comparison see Table~\ref{parameters}). Since the charge ordering as well as the 
charge transport properties strongly depend on the energies described by these parameters, a 
more exact determination of their values could lead to a better understanding of the electronic 
properties of \navo.

\par
In this contribution, we present a joint experimental and theoretical investigation of the electronic 
excitations and their momentum dependence in \navo\, measured using high resolution electron 
energy-loss spectroscopy in transmission. The comparison of the data with models based both on 
the LSDA+U formalism, as well as a cluster approach, enable the construction of a consistent 
theoretical description of the electronic structure of \navo\, and result in the determination of a double-checked parameter set.
The paper is organized as follows. In Sec. II the experimental and theoretical methods are introduced.
Section III contains the presentation and discussion of the experimental and theoretical results and is split
into sub-sections dealing with the different aspects of the data. Lastly, Sec. IV is a summary.

\section{Methodology}
\label{method}

\subsection{Experiment}

\subsubsection{Samples}
\label{samp}

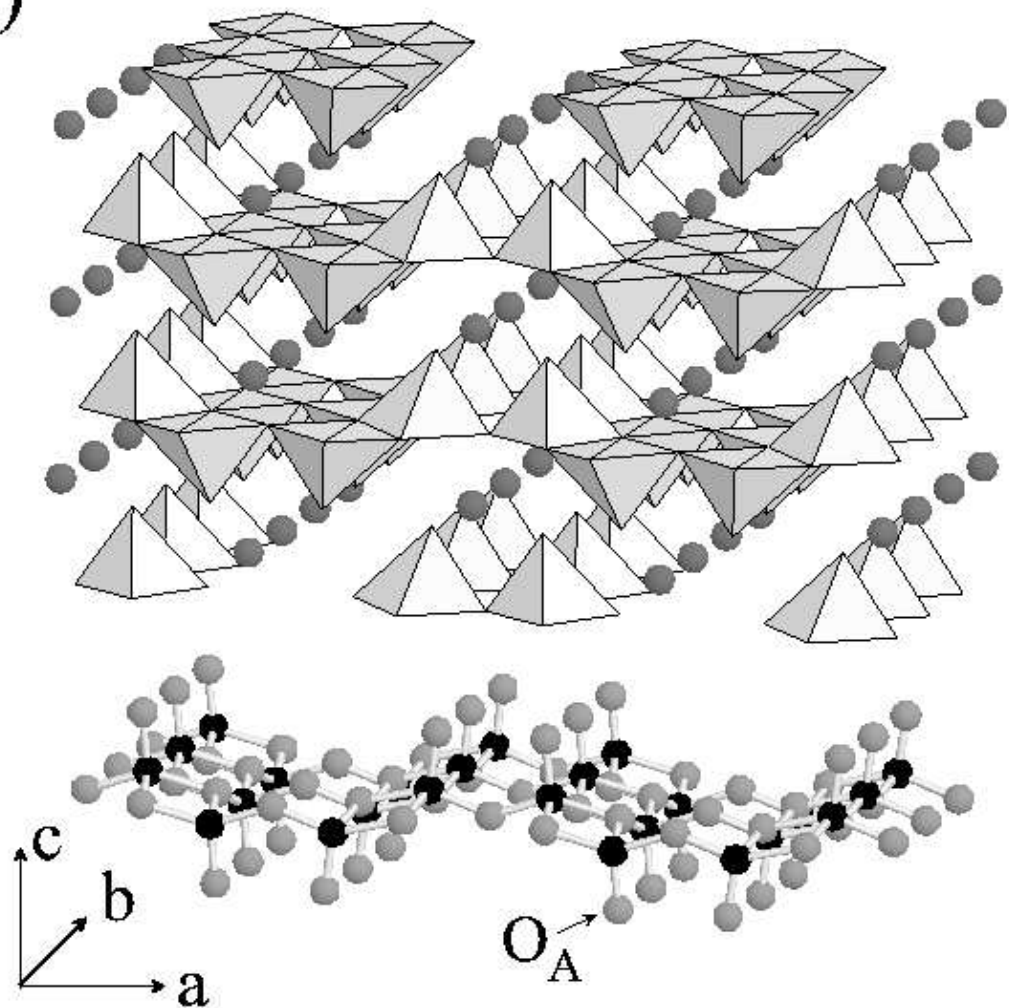
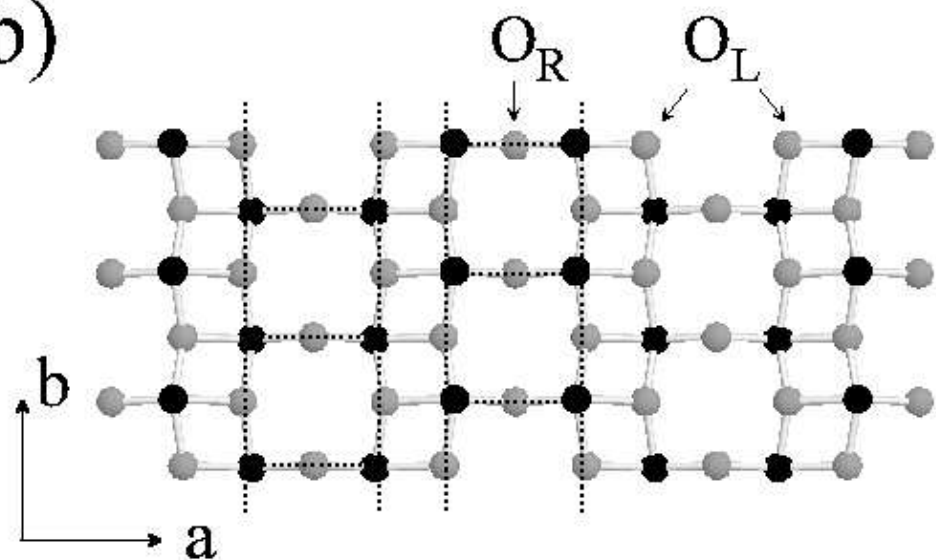
\begin{figure}
\caption{Crystal structure of \navo. Vanadium atoms are shown as black spheres, oxygen atoms 
as grey spheres. (a) Layers of distorted VO$_5$ pyramids (upper part) seperated by sodium ions 
(dark grey spheres) and the atomic ordering of a single vanadium oxide layer (lower part). (b) 
Projection of a single VO-layer onto the ab-plane, excluding the apex oxygens. The ladder 
structure is illustrated by the dashed lines.}
\label{Structure}
\end{figure}

Single crystals of \navo\ were grown from the melt. The detailed procedure is described 
elsewhere.\cite{lohmann97} 
\navo\ crystallizes in an orthorhombic unit cell with lattice constants \mbox{a = 11.318 \AA}, 
\mbox{b = 3.611 \AA} and \mbox{c = 4.797 \AA}.\cite{carpy} The vanadium atoms and their five 
nearest neighbor oxygen atoms build slightly distorted VO$_5$ pyramids which are connected 
by their corners along the {\bf b} direction as well as by their corners and edges along the {\bf a} 
direction, as shown in Fig.~\ref{Structure}a. The resulting pyramid layers in the ab-plane are 
separated by sodium atoms. The weak interactions between the vanadium oxide layers and the sodium 
layers are the reason for the good cleavage behavior of this system along the plane perpendicular to the {\bf c} 
direction. The projection of the vanadium and the oxygen atoms of a single pyramid layer onto the 
$ab$ plane, excluding the oxygen atoms on the apices of the pyramids, delivers the atomic 
configuration as illustrated in Fig.~\ref{Structure}b. This approach emphasizes the ladder structure in 
which adjacent ladders are shifted with respect to each other by half of the lattice parameter b in the {\bf b} direction (see 
dashed lines in Fig.~\ref{Structure}b). 

For the measurements using electron energy-loss spectroscopy in transmission, thin films of about 
1000 \AA\ thickness were cut  from the single crystals  with a diamond knife using an 
ultramicrotome. Because of the good cleavage behavior, the crystallinity remains conserved after 
cutting parallel to the $ab$ plane. The high quality and orientation of the single crystalline samples 
were checked by {\it in~situ} electron diffraction.

\subsubsection{EELS in transmission}
\label{exper}
EELS in transmission with a primary beam energy of 170 keV was performed on free standing 
films at room temperature (for experimental details see Ref.\onlinecite{fink}). The energy and 
momentum transfer (q) resolution were chosen to be 110 meV and 0.05 \AA $^{-1}$ for q $\leq$ 
0.4 \AA $^{-1}$, and 160 meV and 0.06 \AA $^{-1}$ for q $>$ 0.4 \AA $^{-1}$, in order to 
compensate for the decrease 
of the cross section at higher momentum transfer. 

EELS in transmission provides us with the momentum and energy dependent loss function Im(-
1/$\varepsilon$(${\bf q}$, $\omega$)), from which, by means of the Kramers-Kronig relations, the 
real part of the negative inverse dielectric function and thus all the optical properties such as, for 
example, the optical conductivity $\sigma (\omega)$ can be calculated. For small momentum 
transfer only dipole transitions are allowed, and for the limit q = 0 the transition matrix elements are 
the same as in optics. For the Kramers-Kronig analysis the loss spectra closest to the optical limit 
(q = 0.1 \AA$^{-1}$) were used in order to derive the optical conductivity spectra. 

\subsection{Theory}

\subsubsection{Band structure calculations}
\label{bandcalcul}

The band structure of \navo\ was calculated self-consistently using the
scalar-relativistic LMTO method \cite{Andersen} in the atomic sphere
approximation with the combined correction taken into account
(ASA+CC). The von Barth-Hedin parameterization \cite{bh72} was used for the
exchange-correlation potential constructed in the local spin density
approximation. The Brillouin zone (BZ) integrations in the self-consistency
loop were performed using the improved tetrahedron method. \cite{blochl94}

The experimental lattice parameters and atomic positions as determined for the
high temperature phase (Ref.~\onlinecite{schnering}) were used in the
calculations. The
angular momentum expansion of the basis functions included $l=3$ for vanadium
and sodium and $l=2$ for oxygen and the empty spheres.\cite{footnote} The O~3$d$ and
the V~4$f$ states were included in the basis, as they give a significant contribution to
the dipole matrix elements of the optical transitions.

In order to account for the strong electronic correlation at the V sites, we used the LSDA+$U$ method
in the band structure calculations \cite{ldau:ASK93}, which has been 
shown to be very helpful for the description of the electronic structure of
transition metal oxides, in which the 3$d$ orbitals hybridize quite strongly with
the oxygen 2$p$ orbitals (for a review see Ref.\ \onlinecite{ldau:AAL97}). In this
method a Hubbard-like term is added to the LSDA total energy functional:
\begin{equation}
E^{\text{LDA}+U}=E^{\text{LSDA}}+E^{U}-E^{\text{dc}},
\label{eq:ldau}
\end{equation}
where $E^{\text{LSDA}}$ is the LSDA energy functional, $E^{U}$ takes into
account the on-site Coulomb and exchange interactions, and $E^{\text{dc}}$ is necessary to avoid 
the 
double counting of the averaged Coulomb and
exchange interactions already included in $E^{\text{LSDA}}$.\cite{ldau:LAZ95} If non-spherical
contributions to the on-site Coulomb and exchange integrals $U$ and $J$ are
neglected, the $E^{U}$ term can be written as:
\begin{equation}
E^{U}=
\frac{1}{2}\sum_{\sigma,i,j\ne i}(U-J)n_{i\sigma}n_{j\sigma}+
\frac{1}{2}\sum_{\sigma,i,j}U n_{i\sigma}n_{j-\sigma},
\label{eq:eu}
\end{equation}
where $n_{i\sigma}$ is the occupancy of the $i$-th localized orbital with the spin
projection $\sigma$. The localized orbitals used in (\ref{eq:eu}) are
constructed in such a way that they diagonalize the charge density matrix
$n^{\sigma}_{ij}$.
Then, the expression for the orbital dependent one-electron potential corresponding to 
(\ref{eq:ldau}) is given by:
\begin{equation}
V^{\text{LDA}+U}_{\sigma}=V^{\text{LSDA}}_{\sigma}+
\sum_{i} (U-J)(\frac{1}{2}-n_{i\sigma}) |i\sigma><i\sigma|
\label{eq:vldau}
\end{equation}
where $|i\sigma><i\sigma|$ is the projector onto the localized orbital.

The absorptive part of the optical conductivity tensor components was computed
from the LMTO energy bands and eigenvectors on the basis of the
linear-response expressions,\cite{kubo57,maks88,rmme93} whereas the dispersive
part was obtained via the Kramers-Kronig transformation. The finite-lifetime
effects and the experimental resolution were simulated by broadening the
calculated dielectric tensor spectra with a Lorentzian width of 0.2~eV.

As mentioned in the introduction, 
\navo\ can be considered as composed of linear chains of V-O-V rungs containing a single $d$-
electron in a molecular orbital-like state with antiferromagnetic alignment along the ladders.
For the LSDA+$U$ calculations presented here, initially an antiferromagnetic (AFM) order of each of the 
V spins along the leg of the ladder is assumed.
We have also carried out a brief analysis of the influence of deviations from this AFM order upon the 
calculated DOS and optical conductivities by means of spin-waves. 
The band structure calculations for non-collinear magnetic structures including 
constrained moment directions were performed using the formalism developed in
the Refs.~\onlinecite{sandr} and \onlinecite{magnon}. 
The magnetization direction in the atomic sphere surrounding the site ${\bf
t}+{\bf R}_i$, where ${\bf t}$ is its position in the unit cell and ${\bf R}_i$  is
a real space lattice vector, is defined by two polar angles $\theta_{ti}$ and
$\phi_{ti}$. If $\theta_{ti}$ does not depend on ${\bf R}_i$ and $\phi_{ti}$ is given by
\begin{equation}
\phi_{ti}=\phi_{t}+{\bf u}\cdot{\bf R}_i,
\label{eq:spir}
\end{equation}
($\phi_{t}$ determines the relative orientation of the magnetization
directions for different sublattices), the calculations can be
performed without an enlargement of the unit cell for an arbitrary vector ${\bf
u}$, which defines a spiral magnetic structure in real space. (In the present
paper we adopted the notation ${\bf u}$ instead of the usually used character ${\bf q}$ to
distinguish the vector of a spiral structure from the momentum transfer vector
in the EELS measurements).
The details of the $U$ calculations for spiral magnetic structures will be
given elsewhere.\cite{roland} Here we only mention that in this case the
effective orbital dependent potential is still given by equation
(\ref{eq:vldau}) but with the spin part of a localized orbital $|i\sigma>$
centred at a given atomic site being defined in the local coordinate system
with the $z$ axis parallel to the magnetization direction in the corresponding
atomic sphere.

\subsubsection{Cluster calculations}
\label{clustercal}

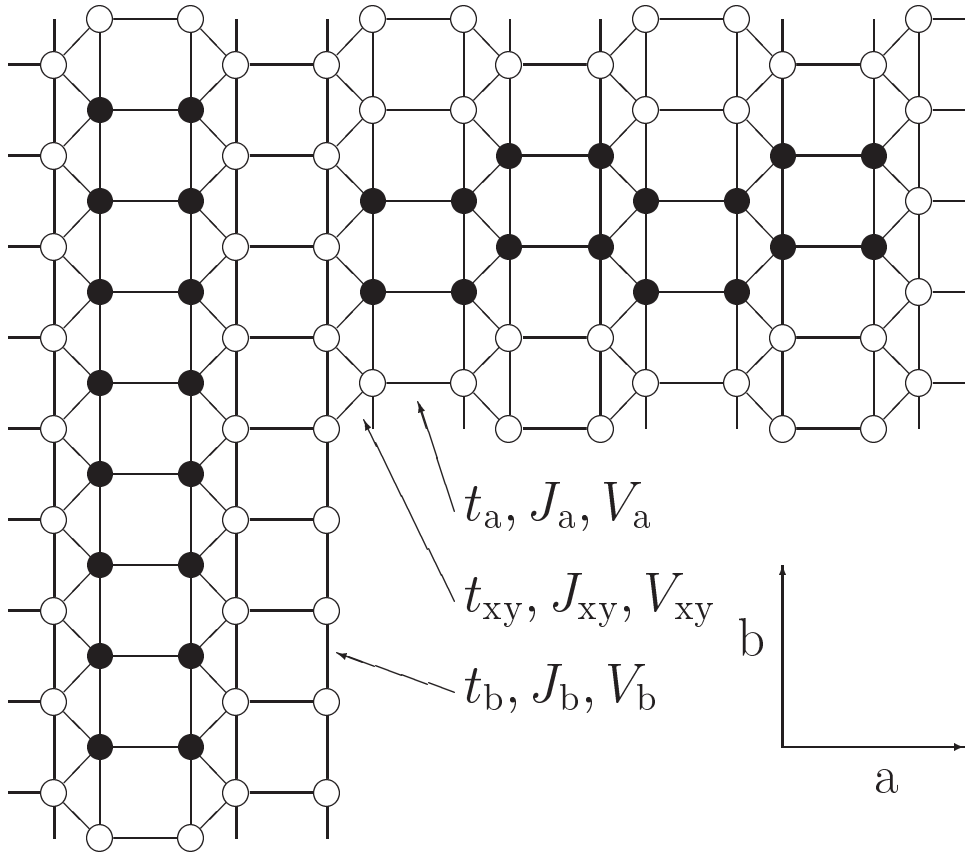
\begin{figure}
  \caption{
    Schematic structure of the $ab$ planes in \navo\ where the 
    circles stand for the vanadium sites. The black sites define the two 
    clusters used in the calculation.
  }
  \label{ebene}
\end{figure}

We have studied the dynamic dielectric response using a quarter-filled $t$-$J$-$V$ model
\begin{eqnarray}
  {\cal H} &=& -\sum_{\langle i,j\rangle,\sigma}t_{ij}
               \left(
                 \hat{c}_{i,\sigma}^{\dagger}\hat{c}_{j,\sigma}+{\rm H.c.}
               \right)
               + \sum_{\langle i,j\rangle}J_{i,j}
               \left(
                 {\rm \bf S}_{i}\cdot{\rm \bf S}_{j}-\frac{1}{4}n_{i}n_{j}
               \right) \nonumber\\
           & & +\sum_{\langle i,j\rangle}V_{i,j}n_{i}n_{j} \label{hamilton}\quad ,
\end{eqnarray}
where $\hat{c}_{i,\sigma}^{\dagger}=c_{i,\sigma}^{\dagger}(1-n_{i,-\sigma})$ are 
constrained electron creation operators, 
$n_{i}=\sum_{\sigma}\hat{c}_{i,\sigma}^{\dagger}\hat{c}_{i,\sigma}$ is the 
occupation-number operator, and ${\rm \bf S}_{i}$ denotes the 
spin-$\frac{1}{2}$ operator at site $i$. The expression $\langle i,j\rangle$ denotes the 
summation over all pairs of nearest neighbors. The hopping parameters 
$t_{ij}$, intersite Coulomb interactions $V_{ij}$, and exchange interactions 
$J_{ij}$ are defined in Fig.~\ref{ebene}.

The loss function measured in EELS experiments is directly proportional to the 
dynamic density-density correlation function.\cite{Schnatterly} By including 
the long-range Coulomb interaction in the model within a random-phase 
approximation (RPA) one finds for the loss function 
\begin{eqnarray}
  L(\omega,{\bf q}) & = & 
    {\rm Im}\left[
      \frac{-1}{1+v_{\bf q}\chi_{\rho}^{0}(\omega,{\bf q})}
    \right],\label{lost}
\end{eqnarray}
where 
\begin{eqnarray}
  \chi_{\rho}^{0}(\omega,{\bf q}) & = &
    \frac{i}{\hbar}\int_{0}^{\infty}dt\langle 
      0|[ \rho_{\bf q}(t), \rho_{-{\bf q}}]|0
    \rangle e^{i\omega t}
\end{eqnarray}
is the  response function at zero temperature for the short-range 
interaction model described by Eq. (\ref{hamilton}). $\chi_{\rho}^{0}$ depends on the energy 
loss $\omega$ and momentum transfer ${\bf q}$. $|0\rangle$ is the ground 
state, 
$\rho_{\bf q}$ denotes the Fourier transform of $n_{i}$, and 
$v_{\bf q}=e^{2}N/(\epsilon_{0}\epsilon_{r}v{\bf q}^{2})$ is the long-range 
Coulomb potential with unit cell volume $v$. 
Furthermore, $N$ is the number of electrons 
per unit cell, $\epsilon_{0}$ is the permittivity, and $\epsilon_{r}$ is the 
real part of the dielectric function.

We evaluated Eq.~(\ref{lost}) by direct diagonalization using the standard 
Lanczos algorithm.\cite{Lin} As this method is limited to small clusters, the 
maximum cluster size in our calculations was restricted to 16 vanadium sites. 
However, calculations for a cluster consisting of two adjacent ladders, each 
containing four rungs, showed strong finite-size effects. We have solved this 
problem by using two different clusters for momentum transfer parallel to 
{\bf a} and {\bf b} direction, as denoted in Fig.~\ref{ebene} by the filled 
circles. In the {\bf a} direction we chose an array of double rungs extending over 
four adjacent ladders, whereas in the {\bf b} direction the cluster consisted of a 
single ladder with eight rungs. The separation into these two clusters is 
justified by the small value of the interladder hopping amplitude 
$t_{xy} = 0.012$~eV (details will be given elsewhere\cite{Huebsch}). 
This value of $t_{xy}$ has also been found by previous LDA calculations\cite{smolinski}, as well as from a 
tight binding fit to our own LDA results.\cite{yaresko} Additional evidence for 
a small $t_{xy}$ follows from the weak magnetic dispersion along the {\bf a} 
direction as observed in neutron scattering experiments.\cite{yosihama}

For the calculation of the loss function we chose open boundary conditions, 
and for momentum transfer parallel to the {\bf a} direction we used renormalized 
intersite Coulomb interactions $\bar{V}_{a}=V_{a}+V_{b}$ and 
$\bar{V}_{b}=2V_{b}$. 
These values follow from a straightforward analysis of 
the influence of adjacent rungs of the same ladder on the electronic states.\cite{Huebsch} Furthermore, 
in the case of open boundary conditions one has to make sure that electrons on 
the edges of the cluster are still embedded in the local Coulomb potential 
that results from a zig-zag charge ordered state. Thus, sites on the edge of the 
cluster for momentum transfer parallel to the {\bf a} direction are assigned an 
additional on-site energy $V_{xy}$. By analogy, sites on the edges of the 
cluster for {\bf q} parallel to the {\bf b} direction need an additional on-site 
energy $V_{b}$, if they are not occupied in a zig-zag charge ordered state.

\section{Results and discussion}
\label{results}

\subsection{Loss function with small momentum transfer}

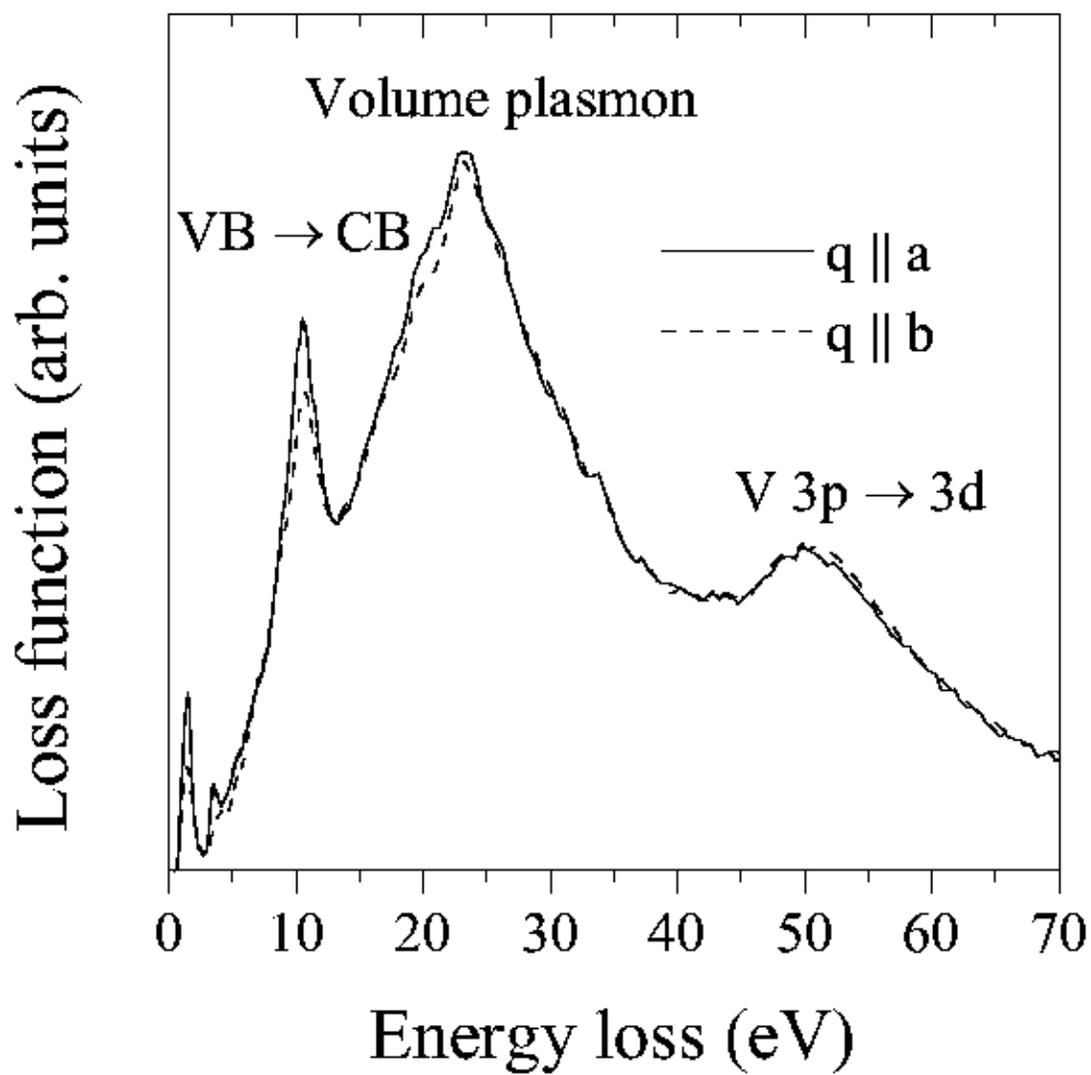
\begin{figure}
\caption{Electron energy-loss spectra of \navo\ measured with momentum transfer $\bf q$~=~0.1~\AA$^{-1}$ 
parallel to the crystallographic $\bf a$ (solid line) and the crystallographic $\bf b$ direction (dashed 
line).}
\label{optlim}
\end{figure}

We have measured the energy dependent loss functions for different momentum transfers {\bf q} 
parallel to the crystallographic {\bf a} and {\bf b} direction in an energy range between 0.5~eV and 
70~eV. Due to the contribution of the elastic line and surface losses it is not possible 
to measure at zero momentum transfer but close to the optical limit (q~=~0.1~\AA$^{-1}$). In 
Fig.~\ref{optlim} the loss functions with momentum transfer {\bf q}~=~0.1~\AA$^{-1}$ parallel to 
the {\bf a} (solid line) and the {\bf b} direction (dashed line) are shown. Both spectra are dominated 
by a broad feature at around 23~eV which represents the volume plasmon - a collective excitation 
of the valence electrons. While the feature at 50~eV results from local excitations between the 
V~3$p$ and the V~3$d$ states, the features emerging below the volume plasmon energy can be 
assigned to transitions of the valence electrons into the conduction band. As Fig.~\ref{optlim} 
shows, an anisotropic behavior is only visible in the low energy regime ($\leq$~22~eV). 

In order to compare the spectra with results derived from our LSDA+$U$ calculations, we have 
calculated the optical conductivity, $\sigma$, by performing a Kramers-Kronig analysis (KKA). 
The comparison with the optical conductivity of the structurally related V$_2$O$_5$ (d$^0$ 
configuration) obtained with the same method,\cite{atzkern} shows that above $\sim$5~eV the 
spectra of both compounds look very similar (not shown). In Ref. \onlinecite{atzkern}, the 
anisotropy of the loss function and of the optical conductivity of V$_2$O$_5$ are discussed in 
detail. As a result, the features above 5~eV can be assigned to transitions from occupied bands 
with mainly oxygen character into unoccupied bands which are dominated by the V~3$d$ states. 
From the resemblance of the spectra and the related crystal structures of both compounds we 
conclude that the same holds for \navo. 
Thus, in this work we focus on the energy range below 5~eV where the loss functions and, 
consequently, the optical conductivities of \navo\ differ strongly from those of V$_2$O$_5$.

\subsection{DOS}
\label{lda}

In comparison to V$_2$O$_5$, which has a similar network of linked VO$_5$ pyramids,
the Na atom in \navo\ donates an additional electron to the valence band. Since in V$_2$O$_5$
the bands formed by the O 2$p$ states are essentially completely filled and the energetically lowest lying 
unoccupied bands are formed by the V~3$d_{xy}$ states, \cite{atzkern} the additional electron in 
\navo\ goes into the rather localized V~3$d_{xy}$ states. These states determine, to a great extent, the 
fascinating properties of this compound.
Due to the localized nature of the V~3$d$ electrons, the Coulomb correlations between them
are rather strong and, as a consequence, conventional L(S)DA calculations fail
to describe properly the electronic structure of \navo. In the unit cell 
corresponding to the centrosymmetric $Pmmn$ structure of the high temperature
phase, a non-magnetic metallic solution is obtained, in contradiction to the
experimental data. 
One way to overcome this discrepancy is to introduce an antiferromagnetic (AFM) order of V magnetic moments
along the {\bf b} direction.
This leads to a doubling of the unit cell and the 
opening of an energy gap in the LSDA band structure.
However, the value of the energy gap is still strongly
underestimated.
It should, however, be borne in mind there is no experimental evidence for the existence
of AFM order in \navo\ even at low temperatures.

An efficient way to take the strong correlations in the V~3$d$ shell into account 
is to use the LSDA+$U$ approach. However, this method is usually applied to
ordered compounds and thus an additional assumption has to be made in order to study
the electronic structure and the optical properties of the high temperature phase
of \navo. Since, according to the experimental data, there is no charge ordering
above $T_c$ and all V sites are equivalent,\cite{ohama97} we have assumed that the effective
one electron potential at each V site is the same and equal to the average of
the LSDA+$U$ potentials for V$^{4+}$ and V$^{5+}$ ions. In other words, the
orbital dependent potential is calculated using the occupation numbers
averaged formally over the V$^{4+}$ and V$^{5+}$ orbital occupation
numbers. On application of $U$, the partially occupied V $d_{xy}$
orbitals are the most affected. The average occupation of the majority-spin $d_{xy}$ 
orbitals is close to 0.5 and their energy position remains unchanged as compared to LDA, whereas 
the unoccupied minority-spin $d_{xy}$ orbitals are shifted upwards by $U/2$. As a
result, an insulating behavior - with the magnetic moment of 0.5 $\mu_B$ per V
atom - or 1 $\mu_B$ per rung - is immediately obtained, which is independent of the
kind of magnetic order which is assumed along the {\bf b} axis. One can say that the on-site Coulomb
repulsion effectively suppresses the occupation of the $d_{xy}$ orbitals with
the opposite spin at each rung.

The effect of the absence, in reality, of long range AFM order in the high temperature
phase of \navo\ on the electronic structure and optical properties was modelled
by performing calculations for spin waves
defined by a vector ${\bf u}=(0,u_y,0)$ with $u_y$ varying from 0.5
(corresponding to AFM order of V magnetic moments along the {\bf b} direction), to 0 at
which all V magnetic moments are ferromagnetically aligned. In these
calculations all $\theta_{t}$ were set to zero, whereas the $\phi_{t}$ were
chosen in such a way that the V ions situated at the same rung had the same
magnetization direction. We found that the relative orientation of
the magnetization at V sites of neighbouring ladders, which is also
determined by $\phi_{t}$, has only minor effect on the calculated band
structure. 

The calculations were iterated to self-consistency for $u_y=0.5$ and then one
iteration was performed for spin waves with other values of $u_y$. The comparison
of the band energies confirmed that the ground state of the system is
antiferromagnetic, in accordance with previous estimates for the exchange
coupling constants between V spins in the {\bf b} direction.

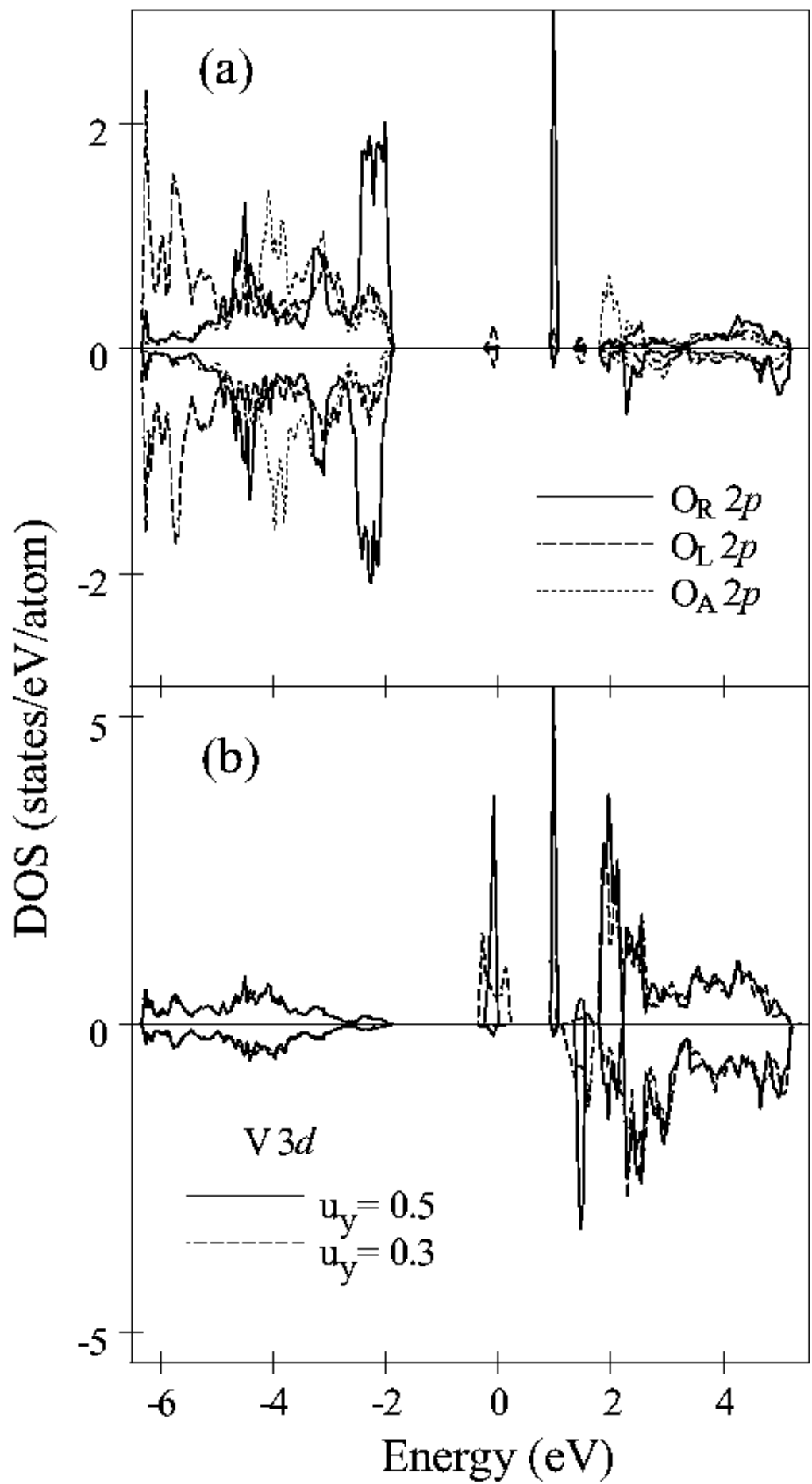
\begin{figure}
\caption{Spin projected densities of (a) the 2$p$ states of the three inequivalent oxygen atoms
and (b) the V~3$d$ states in \navo\ calculated for $U_{\text{eff}}$=3~eV and an AFM
order of V magnetic moments along the {\bf b} direction. For comparison, the densities
of the V~3$d$ states calculated with a vector describing the spiral spin structure of $u_y=0.3$ are also shown by dashed lines in
(b). All energies are given relative to the valence band maximum. In the majority-spin V~3$d$ DOS the two peaks just below the Fermi level and at 1 eV originate from a bonding and antibonding combination of the 3$d_{xy}$ orbitals of the two V atoms on the same rung, respectively. The  corresponding combinations of the minority-spin $d_{xy}$ states are shifted to 1.5 and 2.5~eV by the effective Coulomb repulsion. 
}
\label{DOS}
\end{figure}

The densities of the V~3$d$ and O~2$p$ states, calculated for
$U_{\text{eff}}$~=~3~eV and an AFM order of the V magnetic moments along the {\bf b} direction
are shown in Fig.~\ref{DOS}. 
In Fig.~\ref{DOS}b, the density of V~3$d$
states calculated for $u_y=0.3$ (dashed lines) is also shown.
Here, and in the following, we assume that the
magnetic moments of the V ions situated at the same rung are ferromagnetically
ordered. 
The overall structure of the DOS is similar to that of V$_2$O$_5$.\cite{atzkern}
The states in the energy range from -6.5 to -2~eV are formed mainly by the O~2$p$
states (Fig.~\ref{DOS}a) with a bonding hybridization with the V~3$d$ states. The V~$d_{xz,yz}$ 
states
dominate in the 1.8--3~eV range, while the states with V~$d_{3z^2-r^2}$ and
V~$d_{x^2-y^2}$ character are shifted to higher energies due to the comparatively strong
hybridization with O~2$p$ states and form the upper part ($> 3$~eV) of the conduction
band, shown in Fig.~\ref{DOS}b.
The main differences in the structure of the DOS as compared to V$_2$O$_5$ are
caused by the additional electrons occupying the V~$d_{xy}$ bands which are empty
in the case of V$_2$O$_5$. Since the change in the occupation of the V~$d_{xy}$
states is the main source of the differences between the optical spectra and other 
physical properties of these compounds, in the following we will focus our
attention on the states originating from the V~$d_{xy}$ orbitals.
\par
The narrow peak in the majority-spin V~3$d$ DOS just below the Fermi level (Fig.~\ref{DOS}b)
originates from a combination of the 3$d_{xy}$ orbitals of the two V atoms on the same rung which
is orthogonal to the 2$p_y$ orbital of the O$_{\text{R}}$ atom in the center of the rung. The 
corresponding antibonding
V~3$d_{xy}$--O$_{\text{R}}$~2$p_y$ states are responsible for the DOS peak at
1~eV. In terms of the effective V--V hopping these two peaks result from the bonding and
antibonding combination of the V~3$d_{xy}$ states, respectively, and the energy separation 
between them is
determined by $t_{a}$~=~0.38~eV acting across the rung. 
Here, for the effective V--V hopping terms we use the values obtained from a tight
binding fit to the LDA band structure calculated with the LMTO method,
\cite{yaresko} which were found to be very close to the hopping parameters determined
in Ref.~\onlinecite{smolinski}. 
The minority spin $d_{\downarrow xy}$ states of bonding character are shifted to higher
energy by the effective Coulomb repulsion and lie above the antibonding
$d_{\uparrow xy}$ peak. While the energy of the latter is governed by
$t_{a}$ and thus does not depend on $U_{\text{eff}}$, the relative position of
the minority spin $d_{\downarrow xy}$ states does depend on the exact value of $U_{\text{eff}}$. Finally, the peak
at $\sim$2.5~eV arising from the antibonding $d_{\downarrow xy}$ states is
almost lost in the contributions from the other states of the conduction band, mainly formed by the 
remaining V~3$d$ states hybridized with O~2$p$ states.

It should be noted that for both spin directions the width of the bonding
$d_{xy}$ peaks in the LSDA+$U$ calculations is much smaller than 0.7~eV, the value obtained from LDA calculations,
because the relatively strong hybridization along the {\bf b} direction governed by
$t_{b}$~=~0.17~eV is suppressed due to the AFM order. However, as $u_y$ -
which defines the magnetic structure along the {\bf b} direction - decreases, the width
of the peaks increases and reaches the LDA value when the V magnetic moments are
ordered ferromagnetically. As for the antibonding $d_{xy}$ states, the
dispersion of the corresponding bands along the $\Gamma$--$Y$ direction is much
weaker already in LDA \cite{yaresko} and the width of their DOS does not
depend on $u_y$.

\subsection{Optical conductivity}
\label{optcon}

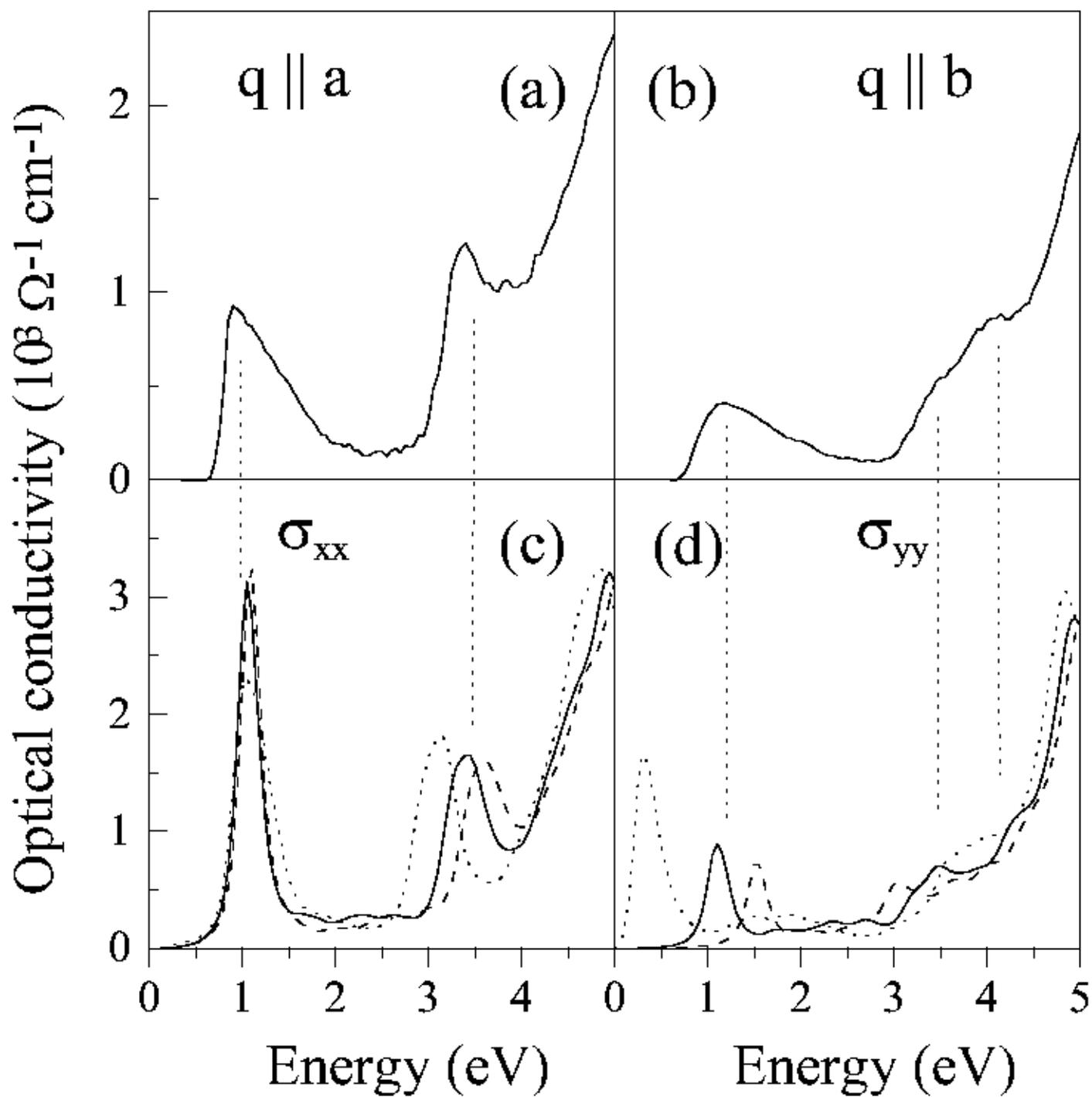
\begin{figure}
\caption{The optical conductivity of \navo.
(a) $\sigma_{xx}$ and (b) $\sigma_{yy}$ from the EELS experiment.
(c) and (d) show the same quantities from the LSDA and LSDA+$U$ band-structure calculations with
$U_{\text{eff}}$~=~0 (dotted line), 2 (solid line) and 3~eV (dashed line).
The theoretical curves are broadened with a Lorentzian of width 0.2~eV.
The vertical dotted lines illustrate the correspondence of the peak positions.}
\label{Th+exp}
\end{figure}

In order to verify our theoretical results we turn our attention to the optical data. In 
Fig.~\ref{Th+exp}, the optical conductivity, derived by means of a Kramers-Kronig analysis of the 
EELS spectra measured at low momentum transfer ({\bf q} = 0.1~\AA), aligned either parallel to the 
crystallographic {\bf a} or {\bf b} direction (Figs.~\ref{Th+exp}a and \ref{Th+exp}b) is shown, together 
with $\sigma_{xx}$ and $\sigma_{yy}$ calculated for $u_y=0.5$ using the LSDA 
($U_{\text{eff}}=0$) and LSDA+$U$ with $U_{\text{eff}}=2$ and 3~eV (Figs.~\ref{Th+exp}c and 
\ref{Th+exp}d). Our experimental spectra agree well with the previous results derived from 
reflectivity\cite{damascelli,long} and absorption\cite{golubchik} measurements of \navo.
Comparing $\sigma_{xx}$ with the {\bf q}$\parallel${\bf a} experimental data (Figs.~\ref{Th+exp}a and \ref{Th+exp}c), one can
see that for this polarization the theoretical spectra are only weakly affected by $U$ and thus that the 
LSDA curve agrees well with the experimental one. The
calculated position of the low energy peak in $\sigma_{xx}$ centred at 1~eV
does not depend on $U$, and coincides almost perfectly with the position of the
corresponding experimental maximum. However, the calculations overestimate the
magnitude of the peak and fail to reproduce accurately its asymmetric shape. The peak
corresponding to the experimental feature at 3.4~eV lies at slightly lower
energy in the LSDA calculations and shifts upwards to $\sim$3.7~eV with
increasing $U$. 

For $\sigma_{yy}$ (Figs.~\ref{Th+exp}b and \ref{Th+exp}d), the situation is quite different. In this case the 
LSDA
calculations result in an optical gap of 0.2~eV, which is significantly
smaller than the experimental value (0.7~eV). 
In contrast to $\sigma_{xx}$, the position
of the low energy peak in $\sigma_{yy}$ is sensitive to $U$, and gets close to the
experimental value for $U_{\text{eff}}=3$~eV. A less strong but still noticeable
dependence on $U$ can be found in the energy range between 2.5 and 4~eV.
Comparing the theoretical spectra calculated with different values of
$U_{\text{eff}}$ to the experimental ones, one can conclude that the best
overall agreement between the theory and the experiment is achieved for
$U_{\text{eff}}$ lying in the range between 2 and 3~eV. 
This range for $U_{\text{eff}}$ derived from the LSDA+U
calculations gives an important guideline for the U value taken in the cluster model 
calculations which will be presented later in the paper.

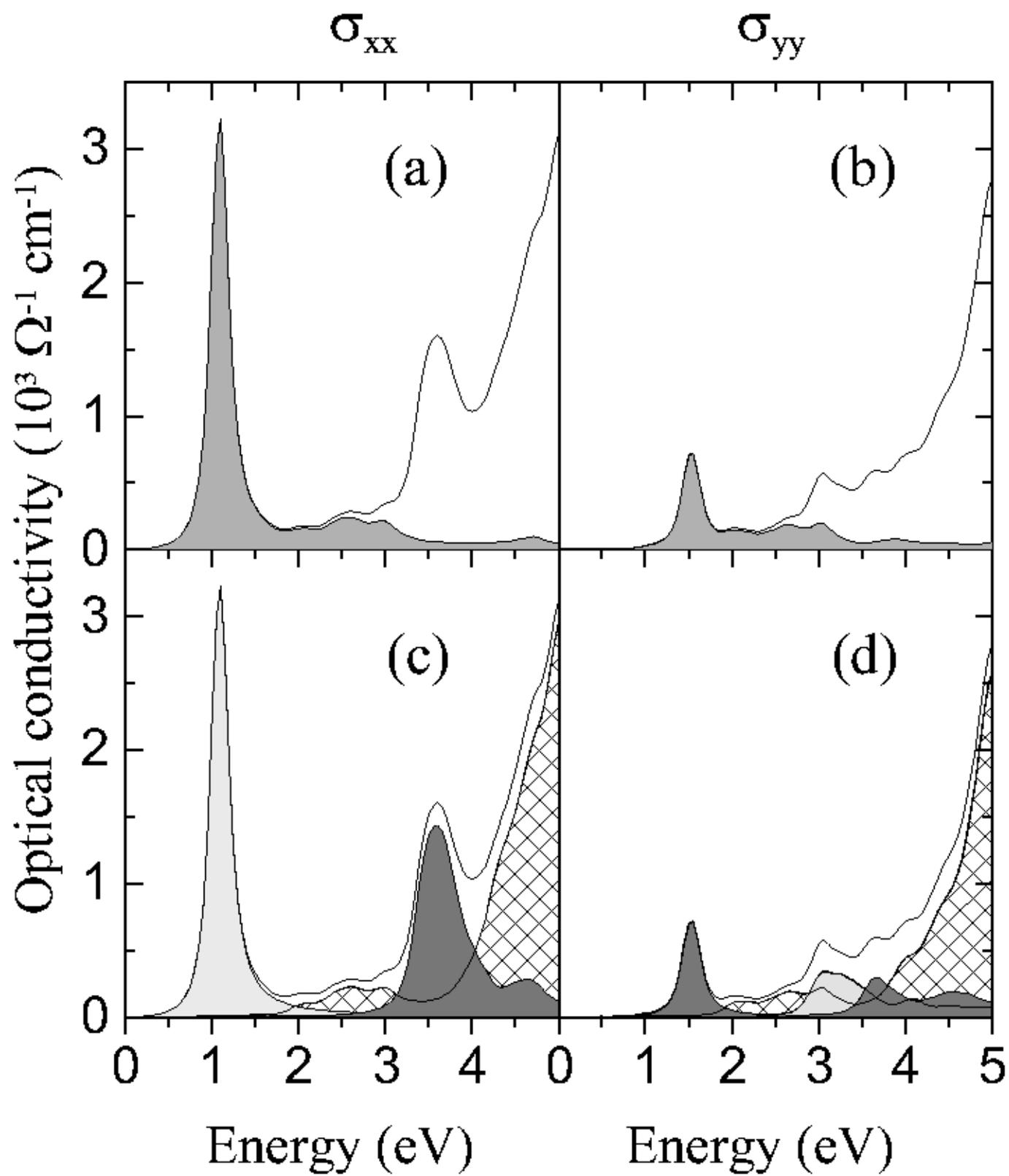
\begin{figure}
\caption{Decomposition of the optical conductivity of \navo\ 
derived from the LSDA+$U$ calculations into contributions arising 
from transitions involving different initial and final states.
In each case panels (a) and (c) show $\sigma_{xx}$ and (b) and (d), $\sigma_{yy}$.
The thin solid lines indicate the total optical conductivities.
In (a) and (b) the grey shading indicates initial states corresponding to occupied V~3$d_{xy}$ bands.
In (c) and (d) light grey (dark grey) shading indicates antibonding V~3$d_{\uparrow xy}$ (bonding V~3$d_{\downarrow xy}$)
final states. The hatched area shows the contribution from interband transitions into higher lying empty V~3$d$ states.}
\label{Sigmas}
\end{figure}

The observed correspondence between the calculated and the experimental spectra allows us to
draw additional information from the detailed analysis of the theoretical
spectra. 
In order to understand better the origins of the features of the optical conductivity
and the reasons for their different dependence on the value of $U$, we analyzed
the contributions to the optical conductivity originating from the interband
transitions between the initial and final states of V~3$d_{xy}$ character.
In Fig.~\ref{Sigmas} we show the results of a decomposition of $\sigma_{xx}$  and 
$\sigma_{yy}$, calculated with $U_{\text{eff}}$~=~3~eV. In each case, the thin solid line
marks the sum of all the contributions - i.e. the total optical conductivity.
Figs.~\ref{Sigmas}a and \ref{Sigmas}c show the calculations for $\sigma_{xx}$, and 
Figs.~\ref{Sigmas}b and \ref{Sigmas}d for $\sigma_{yy}$.

Starting with the top two panels (Figs.~\ref{Sigmas}a and \ref{Sigmas}b), the areas 
shaded grey indicate spectral weight connected with transitions starting from the
occupied 
V~3$d_{xy}$ bands. 
Thus, it is clear that the lowest lying peaks in both $\sigma_{xx}$ and $\sigma_{yy}$ 
are derived from transitions involving V~3$d_{xy}$ {\it initial} states.
The higher lying spectral weight, however, involves transitions starting from 
O~2$p$ states (i.e. the majority of the total $\sigma$ at higher energy is unshaded).

We turn now to an analysis of the final states involved (see Figs.~\ref{Sigmas}c and \ref{Sigmas}d).
In the lower panels of the figure we code the various final state characters with different shading:
(i) the antibonding V~3$d_{\uparrow xy}$ final states are depicted as light grey;
(ii) the bonding V~3$d_{\downarrow xy}$ final states are depicted as dark grey
and (iii) the hatched areas show the contributions from interband transitions into higher lying V~3$d$ states.

On comparing the top and bottom panels of Fig.~\ref{Sigmas} we can therefore conclude the following about the 
transitions giving rise to the optical conductivity.
Firstly, the 1 eV peak in $\sigma_{xx}$ is due to a transition between bonding and antibonding hybrids 
involving the V~$d_{xy}$ levels of equal spin direction.
As was mentioned above, the bonding--antibonding splitting of $d_{xy}$ states is
determined only by the hybridization within a rung and is thus {\it not} affected by $U$. 

Secondly, for momentum transfer parallel to the {\bf b} direction, the matrix elements for
the bonding-antibonding transitions are zero. In this case, transitions between bonding 
V~$d_{xy}$ levels with opposite spin become active.
The energy position of these transitions is sensitive to the strength of the on-site Coulomb
interaction, which naturally explains the strong dependence of the position of the low energy peak in $\sigma_{yy}$ on
the value of $U$ observed in Fig.~\ref{Th+exp}.
In the language of a many-body approach, the $d_{\uparrow xy}\rightarrow
d_{\downarrow xy}$ transitions can be thought of as excitations which lead to the 
creation of doubly occupied rungs.\cite{Footnote1} 

\par
Having dealt with the low-lying features of the optical conductivity, we now
turn our attention to the energy range between 3 and 4~eV. In this context,
the lack of grey shaded weight in Figs.~\ref{Sigmas}a and \ref{Sigmas}b illustrates
that the initial states of the transitions responsible for $\sigma$ in this energy range 
have O~2$p$ character. Consideration of Figs.~\ref{Sigmas}c and \ref{Sigmas}d
reveals that the peak in $\sigma_{xx}$ at 3.7~eV and the small hump in $\sigma_{yy}$ at the 
same energy are dominated by transitions into the $d_{\downarrow xy}$ bands (dark grey shaded
spectral weight).
The initial and final states of the transitions giving rise to $\sigma_{xx}$ between 3 and 4 eV
are, in fact, of the same character as those which determine the shape of the
$\sigma_{xx}$ spectrum in V$_2$O$_5$ just above the absorption threshold.\cite{atzkern}
The dipole transitions from the O~2$p$ valence band states into the antibonding $d_{\uparrow xy}$
states are completely suppressed for a polarization parallel to {\bf a} ($\sigma_{xx}$), while for $\sigma_{yy}$ they play an
important role in the formation of the shoulder in this energy range (shaded light grey
in Fig.~\ref{Sigmas}d). 
\par
Finally, we mention that transitions from the occupied $d_{xy}$ bands to that 
part of the conduction band originating from the V~$d_{xz,yz}$ states give rise to
the non-vanishing intensity in the calculated spectra in the energy range between the 
main features (i.e. between 1.5 and 3 eV). The experimental $\sigma$'s (see Figs.~\ref{Th+exp}a and \ref{Th+exp}b)
confirm the accuracy of the theoretically predicted optical conductivity in this regard.

\par
In order to summarize the detailed information won in this section as regards the character of the states
involved in the transitions giving rise to both $\sigma_{xx}$ and $\sigma_{yy}$ in a concise form, Table~\ref{dec}  
shows a breakdown including the respective initial and final states 'behind' each spectral feature.

\par
The level to which the detailed breakdown of the calculated optical conductivity has been carried out in
Fig.~\ref{Sigmas} is justified by the fact that the theoretical curves (for example the solid line) in Fig.~\ref{Th+exp}
are able not only to reproduce the main features in $\sigma_{xx}$ and $\sigma_{yy}$ quite well, but also the
weaker, double peak structure in $\sigma_{yy}$ in the 3-4 eV range in our experimental data (see Fig.~\ref{Th+exp}b). 
We note that the spectra derived from optical measurements did not resolve this double-peak feature.\cite{damascelli}

\par
Despite the quite good agreement between the calculated and measured optical conductivity on the qualitative level,
the absolute magnitude and strongly asymmetric shape of the low energy peaks in $\sigma_{xx}$ and $\sigma_{yy}$ are
not accurately reproduced by the calculations (see Fig.~\ref{Th+exp}).
We point out two possible sources for these discrepancies: (i) an oversimplified treatment of the charge fluctuations in the high
temperature phase using an averaged one-electron potential 
and (ii) the assumption that the V magnetic moments are antiferromagnetically aligned along the {\bf b} 
direction.
Dealing firstly with the issue of the charge fluctuations, we note that a more rigorous treatment of the charge disorder on
the V sublattice of \navo\ is beyond the capabilities of the LSDA+$U$ approach
in its present formulation. We therefore return to this question later in the paper in the context of the cluster calculations 
which will be described in detail in the next but one sub-section. 
As regards the influence of deviations from long range AFM order on the optical spectra, we point out that the strength of such effects
can be estimated by performing calculations for a spin-wave structure along the b-axis. 

\par
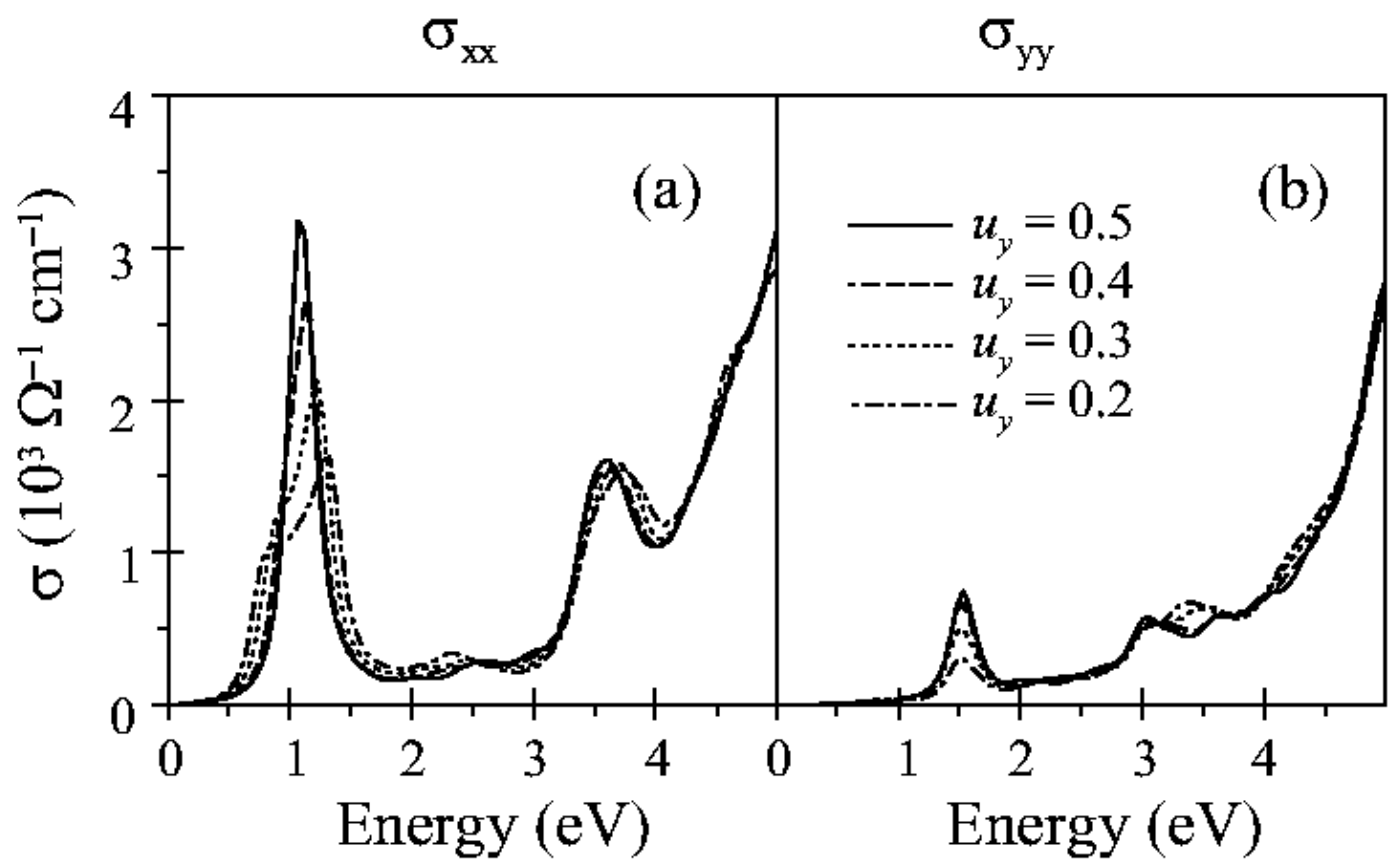
\begin{figure}
\caption{Optical conductivity of \navo\ calculated for $U_{\text{eff}}$~=~3~eV and
different values of $u_y$, which
defines the magnetic structure along the {\bf b} direction (for details see text).}
\label{spirals}
\end{figure}

The optical conductivity spectra calculated for $U_{\text{eff}}$~=~3~eV and
for different values of the vector describing the spiral magnetic structure in real space, $u_y$, are
shown in Fig.~\ref{spirals}.
Since the change of the magnetic structure affects the bands formed from the V~$d_{xy}$ bonding
states most strongly (see Fig.~\ref{DOS}), the spectral features originating from the
transitions in which these bands are involved demonstrate the strongest dependence on
$u_y$.
In particular, the change in shape and intensity of the
peak at 1~eV in $\sigma_{xx}$ (Fig.~\ref{spirals}a) is the most noticeable result of including 
spiral spin structures along the b-axis. 
As can be seen from the figure, as $u_y$ decreases, the peak becomes less
intensive and its maximum shifts to higher energy. Since the final states for
the corresponding transitions (i.e. the antibonding V~$d_{xy}$ states) remain very
narrow independent of the nature of the magnetic order along the {\bf b} direction, the increase in
the width of the peak reflects the change of the density of the occupied V~$d_{xy}$ states (i.e the initial states).
The shape of the low energy peak in $\sigma_{yy}$ (Fig.~\ref{spirals}b) does not change with 
increasing $u_y$ but its intensity rapidly decreases as $u_y$ decreases, and vanishes completely for $u_y=0$
at which point the V magnetic moments order ferromagnetically.
The drop of intensity of the low lying feature in $\sigma_{yy}$ is caused by the decrease of the 
weight of majority-spin states in the final state wave functions.
For $u_y=0$ the initial and final states are formed by pure majority and minority-spin states, respectively, and the
corresponding transitions are forbidden (since the relativistic effects were
not included in the calculations).
The spectral features in the energy range between 3 and 4~eV are also influenced by $u_y$, but the changes are less
pronounced than for the low lying features.
\par
The comparison of the calculated optical conductivity with inclusion of the spin spiral structures (Fig.~\ref{spirals})
with the experimental spectra (Figs.~\ref{Th+exp}a and \ref{Th+exp}b) indicates that an improved agreement 
between the theoretical and experimental magnitude of the low energy peaks of $\sigma_{xx}$ and
$\sigma_{yy}$ can be achieved if deviations from pure AFM order along the {\bf b}-axis are taken into account.
However, even including spin-waves, the LSDA+U results are unable to fully account for the asymmetric shape of the
low energy peaks in the experimental optical conductivities.

To try to overcome this weakness, we have also calculated the optical conductivity for different
relative orientations of the magnetization at neighbouring ladders (not shown).
The changes of the calculated spectra were found to be very small, which is in keeping with the 
size of the inter-ladder hopping term ($t_{xy} =0.012$~eV) which couples V ions
belonging to neighboring ladders.

To bring this section on the data from band structure calculations to a close, the LSDA+$U$ results 
regarding the optical properties of \navo\ can be
summarized as follows.
\par(a) The optical conductivities $\sigma_{xx}$ and $\sigma_{yy}$ calculated with
$U_{\text{eff}}$ in the range between 2 and 3~eV are able to reproduce the main features of
the experimental spectra, as well as the observed optical anisotropy quite well.
This fact indicates the relevance and accuracy of the parameters determined both from 
the fit to the band structure (the transfer integrals, or t's) or from the comparison with the 
experiment ($U$).
\par
(b) The analysis of the contributions to $\sigma$ coming from interband
transitions with different initial and final states allows us to determine the
electronic states responsible for the formation of the peaks in the experimental optical
conductivity. Table~\ref{dec} summarizes this information for each spectral feature in 
$\sigma_{xx}$ and $\sigma_{yy}$.
\par
(c) However, the remaining discrepancy between the theory and the experiment as regards the shape of the
low energy peaks (even despite going beyond the AFM approximation for the spin order along {\bf b})
indicates that the subtle details of the electronic structure of the high temperature phase of \navo\
are beyond the approximations inherent to the LDA, and thus are most likely sensitive to effects
more suitably treated within the framework of models in which electronic correlation is dealt with at a
more fundamental level.

\par
Before going on to the $t-J-V$ model cluster calculations presented in the last sub-section of the results and discussion
part of the paper, the next sub-section deals with the first EELS spectra 
of \navo\ recorded beyond the optical limit.
Here we exploit the strength of EELS in transmission as a measure of the bulk optical properties (in this
case of \navo) for finite momenta, thus giving additional insight into the nature of the two-particle
excitations in this complex system. 

\subsection{Momentum dependent loss functions}

In the left panels of Fig.~\ref{lossf}, we show the EELS loss function of \navo\ recorded for different values
of {\bf q}, either parallel to the crystallographic {\bf a} (Fig.~\ref{lossf}a) or {\bf b} (Fig.~\ref{lossf}a) direction.
The spectra are normalized at higher energy
(ca. 7 eV)
where the shape of the spectra does not change either with the absolute value or with the direction of the momentum 
transfer (see also Fig.~\ref{optlim}).
For clarity the spectra are incrementally off-set in the y-direction. 

For {\bf q} parallel to both the {\bf a} and {\bf b} directions the spectra show strong intensity between the spectral onset at 0.7~eV up to 2~eV, as well as additional weaker features located between 3 and 4.5~eV.
For small {\bf q} the corresponding plasmon excitations are related to the same dipole transitions which are responsible for the peaks at
around 1 and 3.5~eV in the optical conductivities whose origin was discussed in detail above.
In the following, we concentrate our attention on the fine structure of the low energy peaks, as well as on the dependence of 
the shape and position of the peaks on the increasing momentum transfer. 
\par
First, we focus on the features below 2~eV. The intensities in the {\bf a} direction are 
about two times larger than those with the same momentum transfer in the {\bf b} direction. For 
{\bf q}$\parallel${\bf a}, the maximum lies at 1.55~eV
compared with 1.45~eV for {\bf q}$\parallel${\bf b}.
At high momentum transfers the fine structure of the feature around 1.5~eV in the {\bf a} direction indicates the presence of transitions into more than a single final state. 
One can distinguish between essentially three features, whereby the intensity of the component with its
maximum at 1.55~eV decreases more rapidly with q than the features at 1.2 and 1.7~eV, the latter pair being visible as weak 
shoulders only for high momentum transfers.
In the {\bf b} direction the low energy feature is also composed of 
more than one component - at least at high {\bf q}.
The energetically lowest lying  feature (at $\sim$~0.9~eV) - which can be seen as a shoulder at q~$\geq$~0.4~\AA$^{-1}$ -
looses intensity only slowly with increasing q. 
The second component, which one can relate to the main intensity maximum (ca. 1.3 eV) decreases more strongly in intensity
at higher q.
Lastly, for q$\geq0.5$~\AA$^{-1}$, a third component is visible at higher energy (ca. 1.7 eV).

\par
Finally, we turn to the features in the loss function at energies above 3~eV, which, as in the data recorded at the low-q limit,
are clearly separated from the lower energy features by a strong drop of spectral weight between 2 and 3~eV.
In the {\bf a} direction a steep increase of spectral weight occurs at 3~eV, which evolves into a maximum at 3.5~eV.
With increasing momentum transfer the peak position of this feature does not change and the intensity decreases slowly.
A peak at the same energy position with a similar shape was observed for the {\bf a} 
as well as for the {\bf b} direction in the loss functions of V$_2$O$_5$, and is assigned to transitions from the O~2$p$ states into 
the lowest unoccupied V~3$d_{xy}$ states.\cite{atzkern}
However, in the EELS spectra of \navo, the shape of the corresponding feature in the {\bf b} direction differs by the presence of a
double-peaked structure with maxima at 3.5~eV and 5.1~eV.
The two peaks merge together with increasing momentum transfer,  and for q~$\geq$~0.7~\AA$^{-1}$ a further feature becomes visible
at $\sim$3.8~eV. 
Since in electron energy-loss spectroscopy the cross section for dipole transitions decreases and that 
for dipole forbidden transitions increases with increasing momentum transfer, the 
latter 3.8 eV feature is most likely related  to monopole or quadrupole transitions.
\par
Having described the experimental loss functions as a function of momentum, we now discuss the results of the cluster model
calculations aimed at their simulation.

\subsection{Calculated loss functions}

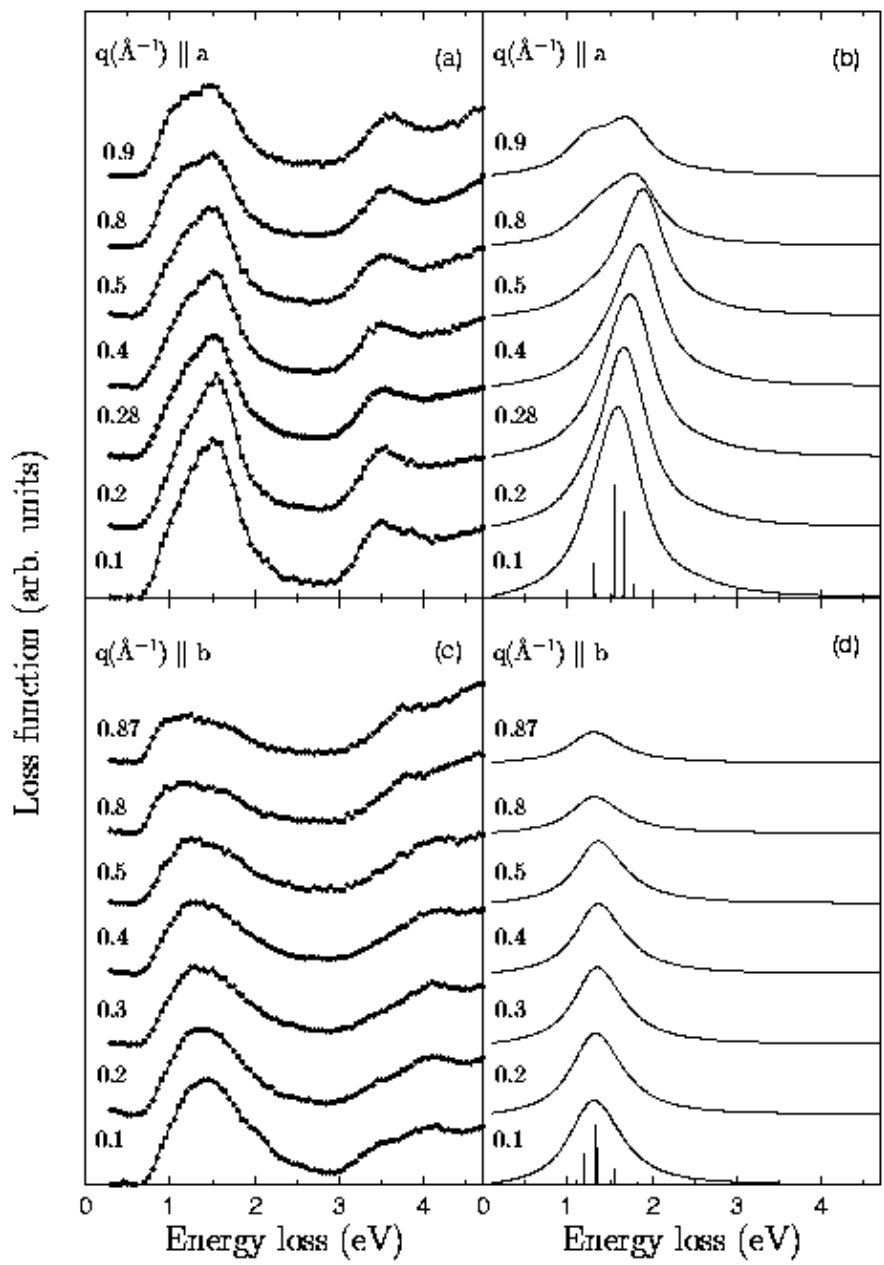
\begin{figure}
  \caption{Panels (a) and (c) show the loss function of \navo\ measured using EELS with momentum transfer 
    aligned along the {\bf a} and {\bf b} directions, respectively.
Panels (b) and (c) show the corresponding calculated loss functions from a quarter filled $t-J-V$ model, which have been broadened with a Gaussian function of width 
0.3~eV. For the q~=~0.1~\AA$^{-1}$ calculated spectra, an unbroadened version has also been 
plotted to enable indentification of the individual plasmons.
For the parameters used in the calculation, see text. 
  }
  \label{lossf}
\end{figure}

In Fig.~\ref{lossf} the results of our cluster calculations (see section 
\ref{clustercal}) are compared with the experimental spectra.\cite{footnote2}
Such cluster calculations require, naturally, input parameters which in turn describe the essentials
of the physics of the system in question.
In this case, for the transfer integrals we take the values $t_{a}=0.38$~eV, $t_{b}=0.17$~eV, $t_{xy}=0.012$~eV from our tight-binding
fit to the band structure calculations mentioned earlier.
These values are identical to those of Ref.~\onlinecite{smolinski}.
A further, vital parameter is the on-site Coulomb repulsion, $U$.
Here we take a value of 2.8 eV - a choice which is guided by the range of $U_{eff} = 2-3eV$ which came out of the LSDA+$U$
calculations presented and discussed above.
\par
At this stage we can have faith in these parameters for two reasons.
Firstly, it is well known that LDA does a good job in describing the charge distribution, and hence the hybridisation / transfer integrals, even
in systems with strong electronic correlation.
Secondly, the good basic agreement between the optical conductivity calculated within LSDA+$U$ with experiment (see Sec.~\ref{optcon})
not only lends support to the $t$'s, but underpins the chosen $U$ value. 
\par
The exchange interactions were parametrized as 
$J_{ij}=4t_{ij}^{2}/U$ and the values of the intersite Coulomb interactions 
$V_{a}=0.8$~eV and $V_{b}=0.6~$eV fulfill the 
condition for the system to be close to a quantum critical point caused by 
charge ordering, as described by Cuoco et al..\cite{cuoco}
The inter-ladder Coulomb interaction $V_{xy}=0.9$~eV has been adjusted within the model 
to obtain the correct peak positions with respect to the experimental loss functions.
To allow a direct comparison with experiment, the calculated data were broadened with a 
Gaussian function of 0.3~eV width.
The unbroadened spectra for ${\bf q}=0.1$~\AA\ are also shown as vertical lines in Figs.~\ref{lossf}b and \ref{lossf}d.
As the cluster model contains only V sites, we only attempt to simulate the loss function in the low
energy range (i.e. for $<$~3eV).
At higher energies - as we know from our combined analysis of the experimental and theoretical optical conductivities (Sec.~\ref{optcon}) - 
excitations from O~2$p$ to V~3$d$ orbitals play a role. Thus this spectral weight cannot be described within the present cluster model.
\par

We find a good agreement between theory and experiment for the low energy part of the loss function for {\bf q}$\parallel${\bf a}, in particular for the 
data recorded at higher q (see Fig.~\ref{lossf}a and \ref{lossf}b).
Specifically, the intensity reduction and increasing width of the experimentally observed structure between $1$ and $2$~eV for q~$>$~0.5~\AA$^{-1}$
is well reproduced in the cluster calculation, a characteristic which eluded our best efforts within the LSDA+$U$ approach. 
For the calculated loss function with {\bf q}$\parallel${\bf b} (Fig.~\ref{lossf}d), the experimental q-dependent reduction in intensity is also well reproduced.
However, in this case the simulation fails to reproduce accurately the large width of the low-energy feature - a shortcoming 
which becomes increasingly apparent for higher q.

\par
In order to enable a discussion of the origin of the individual 
features underlying the calculated spectra, one should note 
that since \navo\ is close to a charge order transition,\cite{cuoco} one can 
describe the nature of the excitations using a zig-zag ordered ground state.
In this context, we mention that excitations with momentum transfer parallel to 
the {\bf a} direction lead to a disturbance of the charge ordering as they involve
hopping of electrons from one side of a rung to the other.
Note that this process can also be interpreted as a
transition from a bonding to an antibonding state of a singly occupied rung, 
in agreement with the results from recent calculations using the Heitler-London
model.\cite{presura}
Since the net inter-ladder Coulomb interaction, $V_{xy}$, remains unchanged after the hopping on the rung in the charge ordered state,
 the excitation energy is dominated by $V_{b}$.
In addition to the one 
electron hopping discussed above, there are also collective processes 
that involve two or three neighboring rungs on different ladders.

\par
As the unbroadened theoretical results presented in  Fig.~\ref{lossf}b show, the theoretical spectra consist of three main features,
each of which can be attributed to a density oscillation (plasmon) with energies between $1$ and $2$~eV. 
For {\bf q}$\parallel${\bf a}, the spectral weight shifts from the plasmon at $1.6$~eV to the one at $1.7$~eV 
with increasing ${\bf q}$, which makes it appear as if the low lying feature as a whole disperses to higher energy,
which is not, in fact, the case. 
The three excitations differ in their degree of delocalization. 
The excitation at $1.3$~eV is related to rather delocalized transitions involving more 
than one electron in the cluster.
The higher lying plasmons, however, result from more localized processes. 

For momentum transfer parallel to the {\bf b} direction, the excitations 
contributing to the low lying spectral weight (see 
Fig.~\ref{lossf}d) can be interpreted as transitions 
from a state in which two rungs are singly occupied to one in which one rung is 
empty and the other doubly occupied.
Due to the 
finite hopping amplitude $t_b=0.17$~eV, the unoccupied and the occupied rung 
are able to move along the ladder independently from each other. Thus, many 
more final states are available in the {\bf b} direction than in the {\bf a} direction.
The energies of the transitions into these final states depend on the distance between the 
unoccupied and the doubly occupied rung. 
However, since we can observe only localized excitations by the direct 
diagonalization of small clusters, we are unable to capture fully the dynamics
of these excitations and thus we miss some spectral weight (in the form of spectral breadth) in the theoretical spectra (Fig.~\ref{lossf}d) 
compared to the experimental data 
(Fig.~\ref{lossf}c).
This explanation is supported by a further 
reduction of spectral weight in the cluster calculation when we adopt a four-rung ladder, in place of the eight
rung cluster from which the results are presented here. 

Thus, in contrast to the good description of the strongly 
localized excitations in the {\bf a} direction described earlier, for {\bf q}$\parallel${\bf b} we find that the theory significantly
underestimates the spectral weight, due to finite size effects in the cluster calculation. 
That these differences also persist in the high q regime illustrates that the cluster size 
imposes limitations even in the case where the excitations have shorter wavelength. 
Nevertheless, it should be pointed out that the cluster calculations do provide a qualitative 
description of both position and the q-dependent decrease of intensity of the experimental features.

\par
The optical conductivity can also be calculated using the same, cluster-based method. With the same
model parameters we obtain a qualitative agreement with the optical 
conductivity, either as derived from EELS or optical measurements \cite{damascelli} (not shown).
However - as was the case for the LSDA+$U$ data - the asymmetric line shape in $\sigma$ could not be fully
reproduced in the cluster calculation.
We ascribe this to the delocalized character of some of the final states, which at {\bf q}~=~0 (i.e. at infinite wavelength)
therefore elude our calculations based upon a finite cluster. 
We note here that a similar type of calculation has been carried out recently,\cite{cuoco} and a good agreement between theory 
and the (q~=~0) optical conductivity was found, albeit with a rather high value of the inter-ladder hopping integral
$t_{xy}=0.15$~eV.
The momentum-dependent EELS data presented here allow a further test of the theoretical parameters beyond the
optical limit. We have found that adopting $t_{xy}=0.15$~eV in our $t$-$J$-$V$ model calculations 
yields a significantly worse agreement with the experimental data than does the value of 
$t_{xy}=0.012$~eV, which also comes out of the tight binding fit to the band structure.
Consequently, it appears as if a large value of the interladder hopping is not a requirement for an accurate description of the charge 
excitations in \navo. 
\par
To summarize, one can say that while the small inter-ladder hopping gives a rationale for the quasi one-dimensional 
character of the dynamics of the V~3$d$ electrons; the fact that all three inter-site Coulomb interactions ($V_a$, $V_b$ and $V_{xy}$)
are of similar size is consistent with the two-dimensional nature of the charge ordering occurring at low temperatures in this system.

\section{Summary}
\label{concl}
In conclusion, we have presented a joint experimental and theoretical investigation of the 
optical properties and collective excitations of \navo.
From measurements by means of high resolution
EELS in transmission, we have derived $\sigma_{xx}$ and $\sigma_{yy}$ in the optical limit and 
have also presented the momentum dependence of the loss function with momentum transfers parallel 
to the crystallographic {\bf a} and {\bf b} directions.
The densities of states and optical conductivities were calculated within an LSDA+$U$ framework, 
with the spin order at the V sites either described in terms of an antiferromagnetic or spin-wave-like arrangement,
whereas the momentum dependent loss functions were simulated using a quarter-filled $t$-$J$-$V$ model
based upon a cluster of 16 vanadium sites.

\par
For {\bf q} parallel to both the {\bf a} and {\bf b} directions, the low energy features of the spectra are essentially
dispersionless - which is in keeping with the small band width of the unoccupied states directly above the
chemical potential predicted in the LSDA+$U$ calculations. 
The comparison between the experimental $\sigma$'s (measured in the optical limit) and those derived from
the LSDA+$U$ calculations yielded a good agreement upon adoption of an on-site Coulomb 
interaction $U = 2-3$~eV and an antiferromagnetic ordering of the spins at the V sites in the ladders. 
However, the LDA-based calculations were not able to reproduce the width of the experimentally
observed structures.
\par
Based upon the LSDA+$U$ results, the differences between $\sigma_{xx}$ and $\sigma_{yy}$ could be 
analyzed in detail and each feature could be characterized according to the nature of the initial and final
states involved in the underlying optical transitions.
For example, it could be shown that the optical conductivity between 0.5 and 1.7~eV is dominated by
transitions from the highest occupied band (bonding combination of V~3$d_{xy}$ states) into the 
lowest unoccupied band (antibonding V~3$d_{xy}$-O$_R$~2$p_y$ states).
While for $\sigma_{xx}$ such an excitation occurs on a single rung, in $\sigma_{yy}$ the same 
transition can be considered as a hopping process to one of the adjacent rungs.
The higher lying features in the optical conductivity (above 3~eV) are shown to be related to 
transitions originating in the O~2$p$ valence band manifold. 
\par
Taking the transfer integrals and $U$ value from our LSDA+$U$ data (the accuracy of which is 
checked by the comparison with the experimental $\sigma$'s), the $t$-$J$-$V$ cluster model
was successful in describing the q-dependence of the intensity and, in particular the 
observed width of the lowest energy feature at high q for {\bf q}$\parallel${\bf a}.
As the high q data probe shorter wavelengths, it is clear that the cluster extension in the {\bf a} direction
is sufficient to describe these localized excitations.
The situation is different for the data with {\bf q}$\parallel${\bf b}. Here the cluster calculation manages to reproduce the 
q-dependence of the intensity well, but still fails to correctly account for the width of the experimentally
observed feature at high q, thus signalling the impact of finite size effects, even in a 16 site cluster.

\par
The analysis of the momentum dependence of the loss function allows a more precise determination of
the values of the model parameters. The best agreement is achieved with transfer integrals which mirror those
derived from the band stucture calculation. In particular, the inter-ladder hopping, $t_{xy}$ is found to be 
0.012~eV, thus confirming the quasi one-dimensional nature of the electronic system in \navo.
Finally, the inter-site Coulomb interactions $V_a$, $V_b$ and $V_{xy}$ are found to be 
of similar magnitude.
These interactions then drive the electronic system close to a quantum critical point between a unordered state
at room temperature and a zig-zag ordered state \cite{vojta} observed at low temperatures.

\acknowledgments
We are grateful to the Deutsche Forschungsgemeinschaft for financial support under contract numbers
DFG Fi-439/7-1 and DFG Ho-955/2-3 .

\begin{table}
\begin{tabular}{ldddddddd}
Ref. & $t_a$ & $t_b$ & $t_{xy}$ & U & $V_a$ & $V_b$ & $V_{xy}$ \\ \hline
Smolinski et al.\cite{smolinski} & 0.38 & 0.17 & 0.012 & 2.8 &&&\\
Horsch et al.\cite{horsch} & 0.35 & 0.15 & 0.3 & 4.0 &&&\\
Damascelli et al.\cite{damascelli} & 0.3 & 0.2 & &&&&&\\
Nishimoto et al.\cite{nishimoto98a} & 0.3 & 0.14 & 0.05 & 4.0 &&0.5\\
Popovic et al.\cite{popovic} & &&& 6.82 &&&\\
Cuoco et al.\cite{cuoco} & 0.4 & 0.2 & 0.15 & 4.0 & 0.8 & 0.8 & 0.9\\
Sa et al.\cite{sa} & 0.38 & 0.17 & 0.012 & 2.8 & 0.37 & 0.10 & 0.43\\
this work & 0.38 & 0.17 & 0.012 & 2.8 & 0.8 & 0.6 & 0.9
\end{tabular}
\caption{Literature values for the hopping parameters, $t$, the on-site Coulomb interaction 
$U$ and the inter-site Coulomb interactions $V_\alpha$ compared with the results of this work.}
\label{parameters}
\end{table}

\begin{table}
\begin{tabular}{ddd}
E (eV) & $\sigma_{xx}$ & $\sigma_{yy}$  \\ \hline
1.1 & ${\rm V}\ 3d_{\uparrow xy}\rightarrow {\rm V}\ 3d_{\uparrow xy}$ &  \\
1.5 & & ${\rm V}\ 3d_{\uparrow xy}\rightarrow {\rm V}\ 3d_{\downarrow xy}$  \\
3.1 & & ${\rm O}\ 2p\rightarrow {\rm V}\ 3d_{\uparrow xy}$ \\
3.7 & ${\rm O}\ 2p\rightarrow {\rm V}\ 3d_{\downarrow xy}$ & ${\rm O}\ 2p\rightarrow {\rm V}\ 3d_{\downarrow xy}$ \\
2\ -\ 3 & ${\rm V}\ 3d_{\uparrow xy}\rightarrow {\rm V}\ 3d_{xz,yz}$ & ${\rm V}\ 3d_{\uparrow xy}\rightarrow {\rm V}\ 3d_{xz,yz}$ \\
\end{tabular}
\caption{Energy positions and character of the electronic transitions contributing to the optical conductivities $\sigma_{xx}$ and $\sigma_{yy}$ of \navo\ in the low energy range.}
\label{dec}
\end{table}

\begin{references}
\bibitem{footnote} To diminish the overlap between atomic spheres 12 additional empty
spheres were inserted into the simple orthorhombic unit cell of \navo.
\end{references}
\end{document}